\documentstyle[prd,aps,floats]{revtex}  

\begin{document}  
\preprint{astro-ph/0006020} \draft 
  
\input epsf 
  
\renewcommand{\topfraction}{0.99} \renewcommand{\bottomfraction}{0.99} 
  
\twocolumn[\hsize\textwidth\columnwidth\hsize\csname 
@twocolumnfalse\endcsname 
  
\title{Initial conditions for hybrid inflation} \author{Lu\'{\i}s E. 
  Mendes and Andrew R.~Liddle} \address{Astrophysics Group, The 
  Blackett Laboratory, Imperial College, 
  London SW7 2BZ, United Kingdom\\ and\\ 
  Astronomy Centre, University of Sussex, Brighton BN1 9QJ, United 
  Kingdom (present address)} 
\date{\today} 
\maketitle 
\begin{abstract}  
  In hybrid inflation models, typically only a tiny fraction of 
  possible initial conditions give rise to successful inflation, even 
  if one assumes spatial homogeneity. We analyze some possible solutions to 
  this initial conditions problem, namely assisted hybrid inflation 
  and hybrid inflation on the brane.  While the former is successful 
  in achieving the onset of inflation for a wide range of initial 
  conditions, it lacks sound physical motivation at present.  On the 
  other hand, in the context of the presently much discussed brane 
  cosmology, extra friction terms appear in the Friedmann equation 
  which solve this initial conditions problem in a natural way. 
\end{abstract}   
  
\pacs{PACS numbers: 98.80.Cq \hfill astro-ph/0006020} 
  
\vskip2pc] 
  
%%%%%%%%%%%%%%%%%%%%%%%%%%%%%%%%%%%%%%%%%%%%%%%%%%%%%%%%%%%%%%%%%%%%%%%%  
\section{Introduction}  
  
Particle physics motivated model building of inflation has undergone a 
renaissance in the last few years, with the realization that the 
hybrid inflation model introduces a natural framework within which to 
implement supersymmetry and supergravity-based models of 
inflation~\cite{cop94,linderiot,doubinf2,ber,LR}.  The important new 
ingredient brought to the picture by supergravity is that the 
potential should only be believed for field values below the reduced 
Planck mass,\footnote{The reduced Planck mass is given by $M_{\rm Pl}=m_{\rm 
Pl}/\sqrt{8 \pi} \approx 0.2 m_{\rm Pl}$, where $m_{\rm Pl}$ is the true Planck 
mass $G^{-1/2}$, $G$ being Newton's constant.} whereas previously only the 
constraint that the total 
energy density be below the Planck scale was imposed. The problem with field 
values larger than the reduced Planck mass is that one expects 
non-renormalizable corrections to the potential of the form 
$(\phi/M_{\rm Pl})^n$ with $n > 4$. For large values of the field 
$\phi$ this may destroy the flatness of the potential therefore 
making the onset of inflation more difficult~\cite{LR}; in any event such terms 
will introduce a large uncertainty in the appropriate form of the potential. 
 
In conventional models of inflation where the inflationary epoch ends 
by leaving the slow-roll regime, it is problematic to obtain 
sufficient inflation subject to this condition on the field values. 
In the hybrid inflation model~\cite{lindehyb}, inflation ends via an 
instability triggered by a second field, obviating the need for the 
troublesome fast-rolling phase, provided the vacuum energy part of the 
potential dominates over the $m^2 \phi^2/2$ term.  However 
Tetradis~\cite{tetradis} has shown that in order for inflation to 
start, the fields must be initially located in a very narrow band 
around the valley of the potential in the direction of the inflaton, 
otherwise the fields will quickly oscillate around the bottom of the 
valley, and pass beyond the instability point in the potential without 
inflation, eventually settling in one of the minima of the potential 
along the axis of the second field. 
 
In this paper we consider two scenarios within which the problem of 
initial conditions, assuming homogeneity, may be solved. Each, in 
different ways, contributes an increase in the Hubble parameter before 
the onset of inflation, therefore enhancing the friction term in the 
Klein--Gordon equation both for the inflaton and for the second field. 
These scenarios are assisted hybrid inflation and hybrid inflation on 
the brane. 
  
We do not directly address the important related question of the 
extent to which spatial gradients in the fields may oppose the onset 
of inflation; the initial conditions problem to which we refer exists 
even if the field is assumed homogeneous. It was recently shown by 
Vachaspati and Trodden \cite{VT} that homogeneity on super-horizon 
scales is required for inflation to commence, confirming previous 
results \cite{early,reviewin}, though it remains unclear the extent to 
which departures from homogeneity are allowed. Nevertheless it is 
clear that any difficulties in obtaining inflation from homogeneous 
initial conditions are likely to exacerbate the problems with 
initiating inflation from more realistic inhomogeneous conditions. We 
mention in passing that an attractive route to solving the spatial 
gradient problem may be topological inflation \cite{top}, where the 
field can be forced into the inflating regime by topological 
considerations and survive there while gradients die away; however we 
do not pursue this idea further here. Other aspects of the initial 
conditions problem have been recently discussed by Felder et 
al.~\cite{FKL}.

\section{Initial conditions for hybrid inflation}  
\label{sec:problem}  
  
We assume that the Universe is described by a flat 
Friedmann--Robertson--Walker model with scale factor $a(t)$. We 
consider the original hybrid inflation potential, given by 
\begin{equation}  
  \label{eq:hyb_pot}  
  V(\phi,\sigma) = \frac{1}{4} \lambda \left( \sigma^2 - M^2 \right)^2  
  + \frac{1}{2} m^2 \phi^2 + \frac{1}{2} \lambda^{\prime} \phi^2  
  \sigma^2 \, ,   
\end{equation}  
where $\phi$ is the inflaton and $\sigma$ the field which triggers the 
end of inflation. We impose the restrictions \mbox{$0 < \lambda\, , 
\lambda^{\prime} < 1$}. Although this particular form of the potential 
does not come directly from any of the particle physics motivated 
inflationary models and may therefore be viewed as a toy model, it 
nevertheless shows the same features as more realistic potentials 
generated in the context of supersymmetry and 
supergravity~\cite{cop94,linderiot,doubinf2,ber,LR}. 
  
For $\phi > \phi_{\rm inst} = \sqrt{\lambda/\lambda^{\prime}} \, M$ 
the potential has a local minimum at $\sigma=0$, corresponding to a 
false vacuum, while for $\phi < \phi_{\rm inst}$ the axis $\sigma=0$ 
becomes a local maximum and the potential has two minima 
$\sigma_{\pm}$ at $\phi=0$, $\sigma=\pm M$. In this model inflation 
happens while the inflaton moves along the valley of the potential at 
$\sigma=0$ for $\phi > \phi_{\rm inst}$. After $\phi$ falls below 
$\phi_{\rm inst}$ the fields quickly move towards the minima 
$\sigma_{\pm}$ of the potential. 
  
The Friedmann equation takes the form 
\begin{equation}  
  \label{eq:fried}  
  H^2 = \frac{1}{3 M_{\rm Pl}^2} \left[ \frac{1}{2} \left(  
      \dot{\phi}^2 + \dot{\sigma}^2 \right) + V(\phi,\sigma) \right]  
\, ,   
\end{equation}  
while the equations of motion for the two scalar fields are 
\begin{equation}  
  \label{eq:eq_phi}  
  \ddot{\phi} + 3 H \dot{\phi} + \frac{dV}{d\phi} = 0 \quad ; \quad 
  \label{eq:eq_sigma}  
  \ddot{\sigma} + 3 H \dot{\sigma} + \frac{dV}{d\sigma} = 0 \,.
\end{equation}  
In this model inflation may happen in one of two distinct parameter regimes: in 
the 
first the quadratic term dominates the potential and we have the usual 
chaotic inflation scenario, while in the second the potential is 
dominated by the vacuum energy, $V\approx \lambda M^4/4$ and we have 
false vacuum inflation. The former regime requires $\phi \gg M_{{\rm Pl}}$, so 
we will focus only on the latter. 
  
In order to obtain the correct amplitude for the density 
perturbations, the masses $m$ and $M$ must satisfy the COBE 
normalization for the density perturbations. In the notation of 
Ref.~\cite{LL}, we take $\delta_{\rm H}=1.9 \times 
10^{-5}$ \cite{BW}. In the case of hybrid inflation the COBE 
constraint cannot be solved analytically, but a useful upper bound on 
the masses can be obtained~\cite{cop94}. For false vacuum inflation we 
get 
\begin{eqnarray}  
  \label{eq:normbigm}  
  \frac{\lambda^{1/4} M }{M_{\rm Pl}} & \lesssim & 2 \times  
  10^{-3} \left( \frac{{\lambda^{\prime}}^2}{\lambda}\right)^{-1/4} \,; \\  
    \frac{m}{M_{\rm Pl}} & \lesssim & 8 \times  
  10^{-5} \left( \frac{{\lambda^{\prime}}^2}{\lambda}\right)^{-1/2} \,,  
\end{eqnarray}  
where we assumed that the relevant scales for COBE leave the horizon 
$60$ $e$-foldings before the end of inflation. 
  
Although hybrid inflation is a very popular model, the beginning of inflation 
requires some fine tuning of the initial conditions. As was 
noted by Tetradis~\cite{tetradis}, it is the balance between the two 
timescales which appear in the equation of motion for the inflaton 
which dictates the fate of inflation: the first timescale is 
associated with friction and is given by $t_{\rm fric} \sim H^{-1}$, 
while the second is associated with the oscillations of the 
inflaton and takes the form $t_{\rm osc} \sim 
1/\sqrt{\lambda^{^\prime}} \, \sigma$ (where we have neglected 
factors of $\pi$). If $t_{\rm fric} < t_{\rm osc}$ inflation begins, 
otherwise the inflaton will quickly roll along the potential. The 
condition for inflation to start in this case reads
\begin{equation}  
  \label{eq:infstart}  
  \frac{\sigma}{M_{\rm Pl}} \lesssim \left( \frac{M}{M_{\rm Pl}}  
  \right)^2 \sqrt{\frac{\lambda}{\lambda'}} \;.  
\end{equation}  
In models where $\lambda' \ll \lambda$ no problem arises, but otherwise then 
even for the largest value of $M$ allowed by the COBE constraint, 
Eq.~(\ref{eq:normbigm}), this requires $\sigma < 10^{-6} M_{\rm Pl}$. 
For the allowed values of the masses, this indeed implies fine tuning 
in the initial conditions. We remind the reader that Eq.~(\ref{eq:hyb_pot}) is 
intended to represent generic hybrid inflation models; for example it might be 
representing the effective potential in models with a large loop correction 
already taken into account.
  
\begin{figure}[t]  
\centering \leavevmode \epsfysize=8.0cm \epsfbox{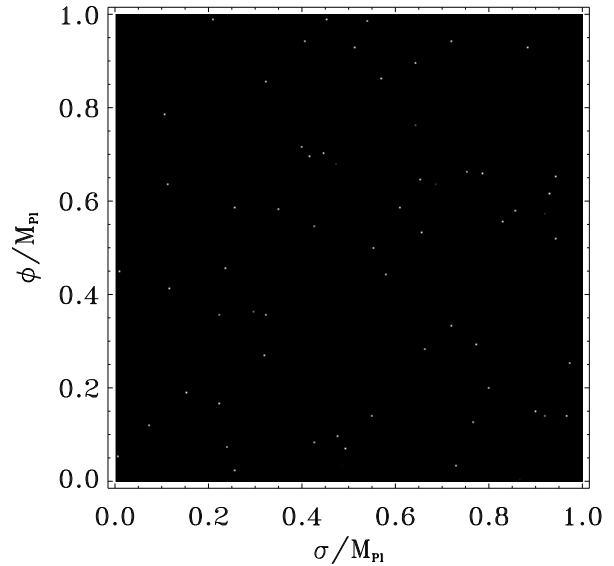} 
\caption[fig1]{\label{fig:hybplot2d} A plot of the  
space $(\phi_{\rm i},\sigma_{\rm i})$ of initial conditions for 
hybrid inflation. White regions correspond to initial conditions 
which give rise to successful inflation ($N_{\rm efold} \geq 70$), 
while black regions correspond to initial conditions which quickly 
lead the fields to the $\sigma_{\pm}$ minima with no successful 
inflation. Although hard to see, the successful region is a very narrow strip 
at the $\sigma = 0$ axis, along with some scattered points off-axis.  These 
plots were obtained with $\lambda=\lambda^{\prime}=1$, $m=5 \times 10^{-6} 
M_{\rm Pl}$ and $M=2 \times 10^{-2} M_{\rm Pl}$, which satisfies the 
COBE normalization.} 
\end{figure}   
 
We have carried out numerical simulations to ascertain which regions of initial 
condition space lead to successful inflation. Fig.~\ref{fig:hybplot2d} shows a 
plot of the space of initial 
conditions with the points which give rise to inflation shown in 
white. Here and in the other 
figures in this paper we used $\dot{\phi}=\dot{\sigma}=0$ which is the most 
favorable case; non-zero values for the derivatives of $\phi$ and 
$\sigma$ will make the fields roll faster along the potential. 
The portion of the space of initial conditions which gives 
rise to inflation is very small; apart from isolated points away from 
the $\sigma = 0$ axis, the space of successful initial conditions is 
restricted to a rather hard to see strip along that axis, as suggested 
by Eq.~(\ref{eq:infstart}). In other words, considerable fine tuning 
in the initial conditions is needed in order to get successful 
inflation. This result is in agreement with the results obtained in 
Ref.~\cite{tetradis}. 

We note that the change in behaviour with varying initial conditions can be very 
sudden. In principle Fig.~\ref{fig:hybplot2d} has a grey scale showing the 
number of $e$-foldings if less than 70, with black corresponding to no inflation 
at all, but in practice the transition between sufficient inflation and no 
inflation is extremely swift. Another implication of this result concerns the 
extent to which the universe
must be in a homogeneous state prior to inflation.  If neighbouring regions have
such vastly different evolution, this may suggest that even mild inhomogeneities
will prevent inflation.  However, such a general statement may not be true; the
final outcome will depend on the dynamical effect of the gradients on the
evolution of the scalar fields.  This has been discussed, albeit in a different
context, by Goldwirth and Piran~\cite{reviewin}, but more detailed investigation 
remains necessary.

\section{Possible resolutions}  
\label{sec:solution}  
  
In this section, we study several mechanisms which may lessen the 
problem of finding viable initial conditions. 
  
\subsection{Including other material}  
  
Since the problem is that inflation is unable to start, there is no 
rationale for considering a universe devoid of matter other than the 
scalar field. The inclusion of extra material provides an additional 
friction to the motion of the scalar field, which helps it towards the 
slow-roll regime. The natural assumption is that we are attempting to 
initiate inflation from within a radiation era, with the scalar field 
coming to dominate. Such radiation will certainly help ease the 
problem of initial conditions; the question is whether it does so 
significantly or not. 
  
In this case the condition for the beginning of inflation mentioned in 
the previous section takes the form 
\begin{equation}  
  \label{eq:eq:inf_start}  
  \frac{\sigma}{M_{\rm Pl}} < \left( \frac{\lambda}{4} M^4 +  
    \rho_{\rm rad}\right)^{1/2} \frac{1}{\sqrt{\lambda^{\prime}} \, M_{\rm  
      Pl}^2} \, .  
\end{equation}  
If we note that the COBE normalization requires $M < 3\times 10^{-3} 
M_{\rm Pl}$, then in order to have inflation even for large values of 
$\sigma$, we would need $\rho_{\rm rad}^{1/4} \sim M_{\rm Pl}$, which 
does not make much sense since we are supposed to be much below the 
Planck scale.  Furthermore, since radiation redshifts as $a^{-4}$, any 
small effect which might be caused by the radiation fluid quickly 
becomes negligible. 
  
Our numerical simulations confirm this simple estimate and show that 
the addition of a radiation fluid to our system does not 
significantly ease the onset of inflation. 
  
\begin{figure*}[t]  
\centering \leavevmode \epsfysize=8.0cm \epsfbox{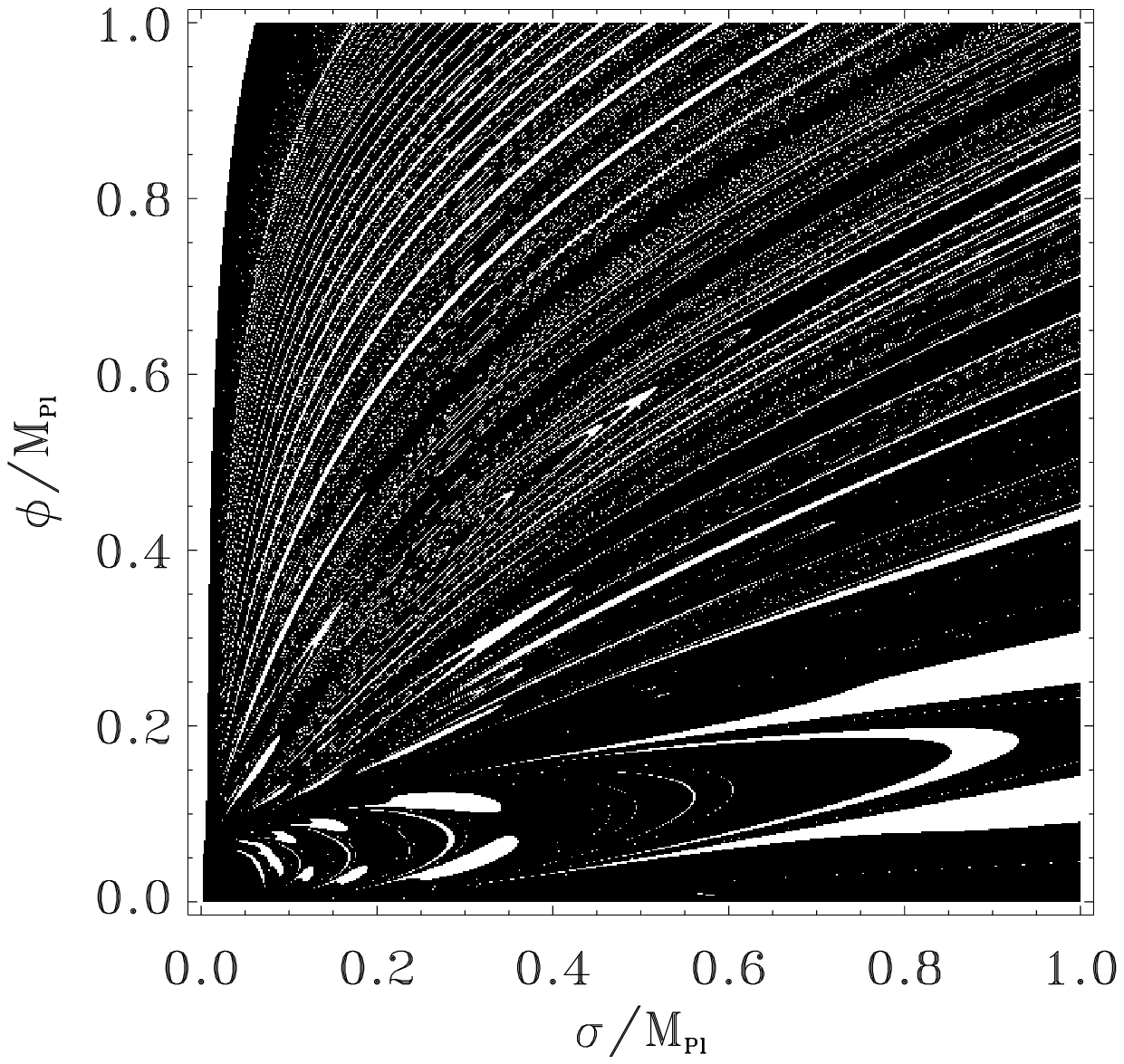} 
\hspace*{0.5cm} \leavevmode \epsfysize=8.0cm \epsfbox{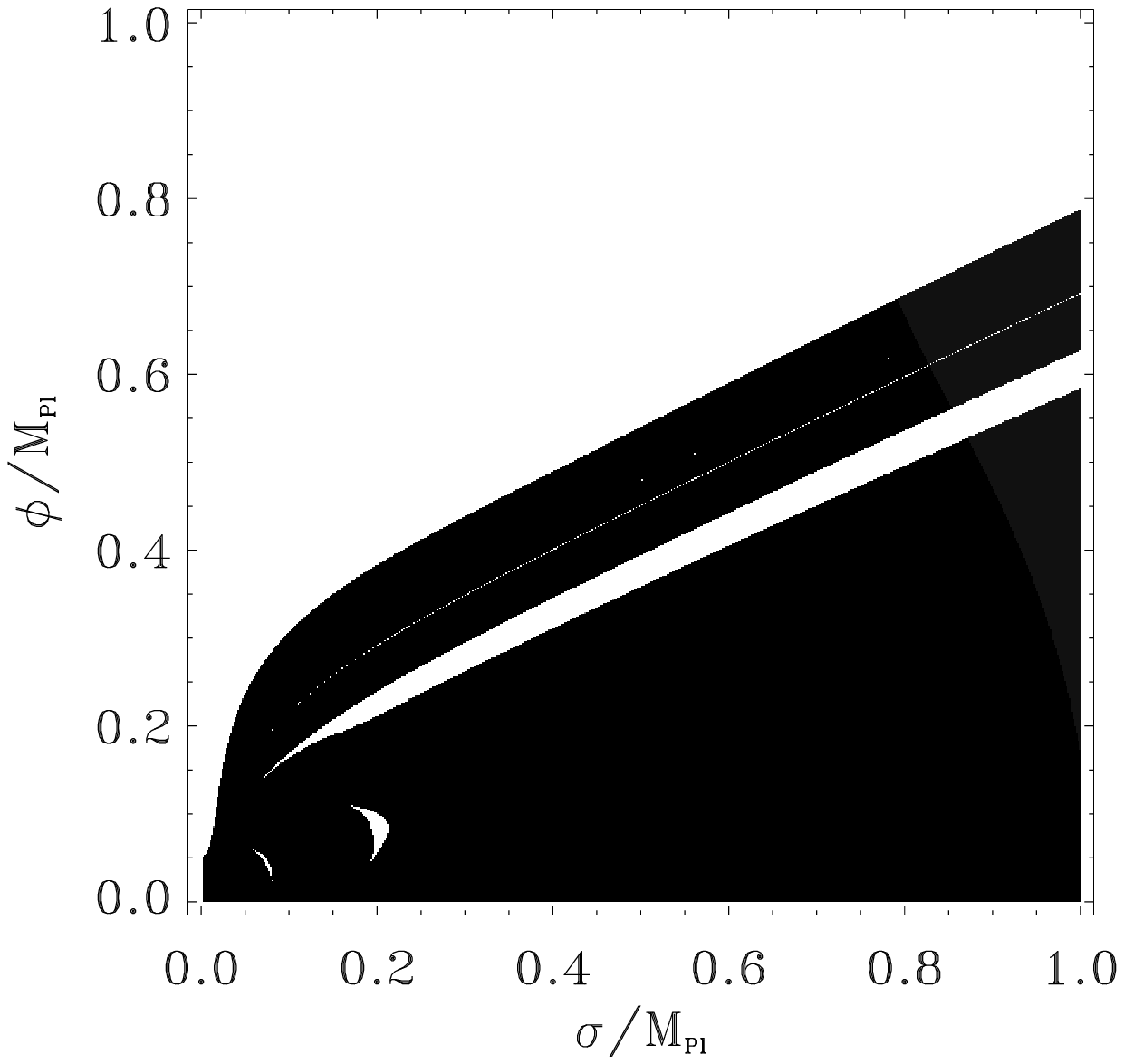} 
\caption[fig1]{\label{fig:hybassplot2d} A plot of the  
space $(\phi_{\rm i},\sigma_{\rm i})$ of initial conditions for 
hybrid assisted inflation with $5$ copies (left panel) and $50$ 
copies (right panel) of the scalar field sector. We assumed copies of the 
original potential with the same masses and initial conditions. As 
before, the white regions indicate successful inflation, and the grey scale 
denotes the number of $e$-foldings with black corresponding to none.} 
\end{figure*}   
  
\subsection{Extra scalar fields}  
  
One resolution to the problem of initial conditions which has been 
discussed is the possibility of an earlier period of inflation which 
sets the initial conditions for the hybrid 
inflation~\cite{doubinf1,doubinf2}.  This does not however appear to 
be a particularly natural evasion, because the question of whether 
suitable initial conditions exist for inflation is simply transported 
onto this new earlier inflation.  However one way in which inflation 
may help is if there are many very similar, or even identical, scalar 
fields as in the case of assisted inflation \cite{assinf}.  Such a 
situation can arise with the compactification of a higher-dimensional 
scalar field, in which many four-dimensional scalar fields arise as 
the Kaluza--Klein tower of states \cite{kanti}. 
  
Assuming we have $n$ copies of the potential, the Friedmann 
equation~(\ref{eq:fried}) is replaced by 
\begin{equation}  
  \label{eq:fried_ass}  
  H^2 = \frac{1}{3 M_{\rm Pl}^2} \sum_{i=1}^n\left[ \frac{1}{2} \left(  
      \dot{\phi_i}^2 + \dot{\sigma_i}^2 \right) +   
    V(\phi_i,\sigma_i)\right] \, ,  
\end{equation}  
where all the copies of the inflaton and the second scalar field still 
obey Eq.~(\ref{eq:eq_phi}). 
  
Fig.~\ref{fig:hybassplot2d} shows the results of our numerical simulations,
where we assumed all the copies of the scalar fields had the same masses and
initial conditions.  With even a moderate number of copies of
the potential, the onset of inflation is now possible in most of the space of
permitted initial conditions.  The situation improves the more copies of the
fields there are. Further simulations we carried out show that 
provided at least one of the sets of fields starts in a region which 
would inflate in the model with only one set of fields, then inflation 
will be able to start. Even if none of the sets of fields is in a 
region which would give rise to inflation in the original model, 
inflation can still start in most cases. 

It is interesting to note that the plots exhibit a fractal-like 
structure with 
new filaments appearing at every scale. Although we have not pursued 
this point, it is not surprising because it is known that 
two-field models of inflation can show chaotic behavior \cite{chaos}. 
The hybrid inflation model can be seen as two coupled harmonic 
oscillators and it is well known that such systems usually exhibit 
chaotic behavior.
 
While successful in alleviating the initial conditions problem, the 
drawback is that we have simply presumed that it is possible to obtain 
large numbers of copies of the hybrid inflation sector. We are not 
aware of any mechanism which is capable of replicating the complicated 
hybrid inflation sector; the Kaluza--Klein implementation of assisted 
inflation \cite{kanti} does not give such a simple outcome in the case 
of coupled fields.

\subsection{Brane cosmology}  
  
The realization~\cite{ark} that we may live on a so-called `brane' 
embedded in a higher-dimensional cosmology has enormous implications 
for cosmology.  One such is the possibility of a modification to the 
Friedmann equation at high energies; Bin\'etruy et al.~\cite{bin} and, 
in a more general sense, Shiromizu et al.~\cite{shiro} have shown that 
the Friedmann equation can acquire a term proportional to the density 
squared (see also Ref.~\cite{perts} for discussion of perturbations). 
This naturally provides additional friction operative at high 
energies. 
  
We will consider here a five-dimensional model similar to the one used 
by Maartens et al.~\cite{maart} in their study of chaotic inflation on 
the brane. In this case the Friedmann equation takes the form 
\begin{equation}  
  \label{eq:fried_brane}  
  H^2 = \frac{\Lambda_4}{3} + \left( \frac{1}{3 M_{\rm Pl}^2} \right)  
  \rho + \left( \frac{4 \pi}{3 M_5^3} \right)^2 \rho^2 + \frac{{\cal E}}{a^4}  
\end{equation}  
where $M_5$ is the fundamental five-dimensional Planck mass, which is 
taken as much smaller than $M_{\rm Pl}$. The effective cosmological 
constant on the brane, $\Lambda_4$, can be expressed in terms of the 
five-dimensional cosmological constant in the bulk, $\Lambda$, the 
tension of the brane $\lambda_{\rm b}$, and the fundamental scale 
$M_5$ as 
\begin{equation}  
  \label{eq:lambda_brane}  
  \Lambda_4 = \frac{4 \pi}{M_5^3} \left( \Lambda + \frac{4 \pi}{3  
      M_5^3} \lambda_{b}^2 \right) \, .  
\end{equation}  
To have a negligible cosmological constant in the early universe 
requires $\Lambda = -4 \pi \lambda_{b}^2/3 M_5^3$. The effective 
four-dimensional Plank mass can be expressed in terms of $\lambda_{b}$ 
and $M_5$ as 
\begin{equation}  
  \label{eq:eff_planck}  
  M_{\rm Pl} = \sqrt{6} \, \left( \frac{M_5^3}{\sqrt{\lambda_{b}}}  
  \right) \, .  
\end{equation}  
As usual, in Eq.~(\ref{eq:fried_brane}), $\rho=\dot{\phi}^2/2 + 
\dot{\sigma}^2/2 + V(\phi,\sigma)$. 
  
The constant ${\cal E}$ in Eq.~(\ref{eq:fried_brane}) describes the 
effect of massive gravitons on the brane and its origin can be traced 
to the projection of the five-dimensional Weyl tensor~\cite{shiro}. We 
will from now on neglect this term. It is clear that the presence of 
the $a^4$ denominator will suppress it during inflation.  Although 
this term may not be negligible at the beginning of inflation, it will 
contribute to the friction term in the equations of motion for the 
scalar fields and can in principle also help alleviate the fine tuning 
in the initial conditions provided ${\cal E}$ is positive. 
  
The equations of motion for the fields $\phi$ and $\sigma$ take the 
usual form, Eq.~(\ref{eq:eq_phi}). Using 
Eqs.~(\ref{eq:lambda_brane}) and~(\ref{eq:eff_planck}) 
in Eq.~(\ref{eq:fried_brane}), and putting $\Lambda_4=0$ and ${\cal E} = 
0$ the Friedmann equation becomes 
\begin{equation}  
  \label{eq:fried_brane_final}  
  H^2 = \frac{1}{3 M_{{\rm Pl}}^2} \rho \left( 1 + \frac{\rho}{2  
      \lambda_{b}} \right)  \,.
\end{equation}  
{}From the previous equation we see that for high energies ($\rho > 
\lambda_{b}$) the quadratic term dominates; later on during the 
radiation epoch, this term will be redshifted by a factor $a^{-8}$ and 
should therefore become negligible. In order that no 
incompatibilities arise with nucleosynthesis the brane tension must 
satisfy $\lambda_{b} \gtrsim 1 \, {\rm MeV}^4$, implying $M_5 > 10 \, 
{\rm TeV}$~\cite{cline}. Here we will mainly be interested in the case 
where the quadratic term in $\rho$ dominates during inflation, that is 
$\lambda_{b} \ll V$. 
  
As in the case of hybrid inflation in the standard four-dimensional 
context, the masses $m$ and $M$ cannot be freely chosen but are 
constrained by the COBE normalization of the spectrum of density 
perturbations. The calculation follows closely that of Ref.~\cite{cop94} for the 
standard four-dimensional case. In our model 
the slow-roll parameters are given by~\cite{maart} 
\begin{equation}  
  \label{eq:eps_brane}  
  \epsilon \simeq 2 M_{\rm Pl}^2 \, \left(  
    \frac{V^{\prime}}{V} \right)^2 \frac{\lambda_{b}}{V} \quad ; \quad  
  \label{eq:eta_brane}  
  \eta \simeq 2 M_{\rm Pl}^2 \, \left(  
    \frac{V^{\prime\prime}}{V} \right) \frac{\lambda_{b}}{V} \, ,   
\end{equation}  
where we assumed $V \gg \lambda_{b}$. The spectrum of density 
perturbations takes the form \cite{LL} 
\begin{equation}  
  \label{eq:pert}  
  \delta_{\rm H}^2 \simeq \frac{1}{600 \pi^2 M_{\rm Pl}^6}  
  \frac{V^3}{{V^{\prime}}^2} \frac{V^3}{\lambda_{b}^3}  \,, 
\end{equation}  
and the number of $e$-foldings is 
\begin{equation}  
  \label{eq:efold}  
  N \simeq -\frac{1}{2 M_{\rm Pl}^2}\int_{\phi_i}^{\phi_f}  
  \frac{V^2}{\lambda_{b} V^{\prime}} d\phi \, .  
\end{equation}  
  
Although there is no general analytical expression for the relation 
between the masses, there are two limits where an analytical 
expression for the masses can be found. When the quadratic term in the 
potential dominates we have chaotic inflation, which is essentially the 
same case as analyzed by Maartens et al.~\cite{maart}. We are interested in the 
case where the false vacuum energy dominates, 
where we have 
\begin{equation}  
  \label{eq:brane_vac_norm}  
  M = 1.4 \times 10^{-1}\left( \lambda\lambda^{\prime} \right)^{-1/6}   
  \eta^{1/3} e^{20 \eta} \lambda_{\rm b}^{1/6} M_{\rm Pl}^{1/3}\,\, ,  
\end{equation}  
where 
\begin{equation}  
  \label{eq:brane_eta}  
  \eta = \frac{3}{\pi^2} \frac{m^2 M_5^6}{M^8}  \,.
\end{equation}  
Using the slow-roll condition $\eta < 1$, we can obtain an upper limit for 
the mass $m$: 
\begin{equation}  
  \label{eq:m_upper_bound}  
  m < 6.2 \times 10^{-5} \left(  
    \frac{\lambda}{{\lambda^{\prime}}^2} \right)^{1/3}  
  \eta^{4/3} e^{80\eta} \lambda_{\rm b}^{1/6} M_{\rm Pl}^{1/3} \, .  
\end{equation}  
We are assuming $\lambda M^4/4 \gg m^2 \phi^2/2$, which imposes a 
further constraint 
\begin{equation}  
  \label{eq:max_eta_vac}  
  \eta^3 e^{240 \eta} \ll 2.3\times 10^6 \,  
    \frac{{\lambda^{\prime}}^2}{\lambda}  \,,
\end{equation}  
which, for $\lambda=\lambda^{\prime}=1$, saturates for $\eta=\eta_{\rm MAX} 
=0.09$. This is exactly the same constraint as in 
the standard cosmological scenario (compare with Eq.~(2.45) 
in Ref.~\cite{cop94}). Using this constraint in 
Eqs.~(\ref{eq:brane_vac_norm}) and~(\ref{eq:m_upper_bound}), the 
vacuum domination approximation gives 
\begin{eqnarray}  
  \label{eq:bigmupper}  
  \lambda^{1/4} M & < & 3.8 \times 10^{-1} \left(  
    \lambda\lambda^{\prime}\right)^{-1/6} \lambda_{\rm b}^{1/6} M_{\rm  
    Pl}^{1/3} \,; \\   
  \label{eq:mupper}  
  m & < & 1.1 \times 10^{-3} \left(  
    \frac{\lambda}{{\lambda^{\prime}}^2} \right)^{1/3} \lambda_{\rm  
    b}^{1/6} M_{\rm Pl}^{1/3} \, \, .  
\end{eqnarray}  
When $\eta \ll 1/20$, the exponential in Eq.~(\ref{eq:brane_vac_norm}) 
can be neglected and we obtain a relation between $M$ and $m$; 
\begin{equation}  
  \label{eq:bim_m}  
  M=0.8 \left( \lambda^5 \lambda^{\prime}\right)^{-1/22}  
  \lambda_{\rm b}^{3/22} m^{2/11} M_{\rm Pl}^{3/11}  
\end{equation}  
  
\begin{figure}[t]  
  \centering \leavevmode \epsfysize=8.0cm 
  \epsfbox{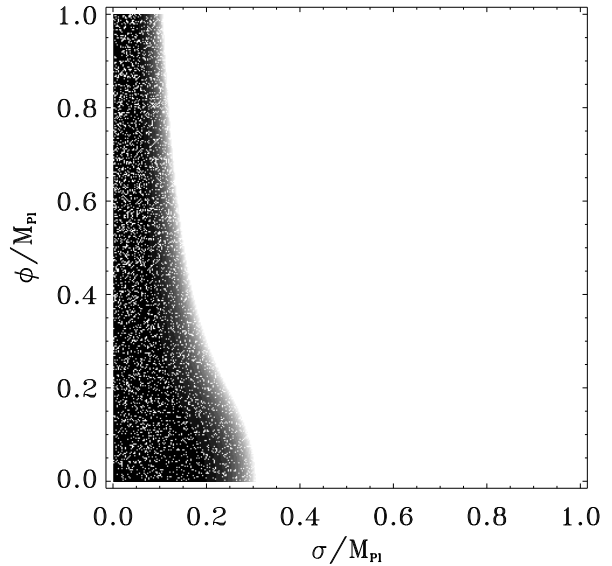} 
\caption[fig2]{\label{fig:hybbraneplot2d} This plot shows the space of  
  initial conditions for hybrid inflation on the brane, with 
  $\lambda_{b}=6 \times 10^{-8} \, M_{{\rm Pl}}^4$, $m=2 \times 10^{-6} \, 
M_{{\rm Pl}}$, and $M=10^{-2} \, M_{{\rm Pl}}$. As before, white indicates 
regions of successful inflation.} 
\end{figure}   
 
Fig.~\ref{fig:hybbraneplot2d} shows our results for hybrid inflation on 
the brane.  The $\rho^2$ term is 
indeed efficient in broadening the region in the space of initial 
conditions where inflation occurs. As we decrease the value of 
$\lambda_{b}$, the onset of inflation becomes even easier as this 
further increases the friction term in the scalar field equations of 
motion. There is also a smoother transition between regions where successful 
inflation 
occurs (at least $70$ $e$-foldings of expansion) and those where 
inflation cannot start. 
  
\section{Conclusions}  
\label{sec:con}  
 
It is debatable whether one should be too concerned that only a 
limited region of initial condition space leads to sufficient hybrid 
inflation in the models we've considered, as one can readily argue, either 
probabilistically or 
anthropically, that the Universe we live in would originate from one 
of these regions. Nevertheless, it is a relevant question as to 
whether or not the problem is mitigated once the scenario is 
generalized beyond a universe containing only the scalar field, and we 
have studied several alternatives. 

The assumption that filling the universe with some other material besides the
scalar fields needed for hybrid inflation would help does not work as one might
expect, mainly because a radiation fluid with a physically reasonable density
will not decrease the friction timescale sufficiently to induce the onset of
inflation.  When we add a large number of copies of the potential to our
original model, in the spirit of assisted inflation, we can in principle obtain
inflation from most of the space of initial conditions.  However, as we pointed
out before it is not obvious how such a number of fields could be introduced in
a physically sensible manner.

Finally, we have studied hybrid inflation in the context of brane 
cosmology. Using a simple model, we were able to show the hybrid 
inflation on the brane does not suffer from the same fine-tuning 
problems in the initial conditions as its standard counterpart. Since 
brane cosmology has become such a popular topic, it is encouraging to know that 
besides being a possible solution for the hierarchy 
problem, it is also possible to have inflation on the brane and, 
furthermore, free from some of the fine-tuning problems that plagued 
conventional inflation.

%%%%%%%%%%%%%%%%%%%%%%%%%%%%%%%%%%%%%%%%%%%%%%%%%%%%%%%%%%%%%%%%%%%%%%%%  
\section*{Acknowledgments}  
L.E.M. is supported by FCT (Portugal) under contract PRAXIS XXI 
BPD/14163/97. We thank Ed Copeland, David Lyth, Anupam Mazumdar and Arttu 
Rajantie 
for useful discussions.  We acknowledge the use of the Starlink 
computer system at the University of Sussex. Part of this work was 
conducted on the SGI Origin platform using COSMOS Consortium 
facilities, funded by HEFCE, PPARC and SGI. 
  
%%%%%%%%%%%%%%%%%%%%%%%%%%%%%%%%%%%%%%%%%%%%%%%%%%%%%%%%%%%%%%%%%%%%%%%%  
   
%%%%%%%%%%%%%%%%%%%%%%%%%%%%%%%%%%%%%%%%%%%%%%%%%%%%%%%%%%%%%%%%%%%%%%%  
\end{document}